\documentclass[showpacs,11pt]{revtex4}
\usepackage{amsmath}
\usepackage{amssymb}
\usepackage{amsfonts}

\begin{document}

\title{Couplings between Chern-Simons gravities and $2p$-branes}
\author{Olivera Mi\v{s}kovi\'{c} }
\email{olivera.miskovic@ucv.cl}
\affiliation{Instituto de F\'{\i}sica, P. Universidad Cat\'{o}lica de Valpara\'{\i}so,
Casilla 4059, Valpara\'{\i}so, Chile.}
\author{Jorge Zanelli}
\email{z@cecs.cl}
\affiliation{Centro de Estudios Cient\'{\i}ficos (CECS), Casilla 1469, Valdivia, Chile.}
\date{\today }

\begin{abstract}
The interaction between Chern-Simons (CS) theories\textbf{\ }and localized
external sources ($2p$-branes) is analyzed. This\textbf{\ }interaction
generalizes the minimal coupling between a point charge ($0$-brane) and a
gauge connection. The external currents that define the $2p$-branes are
covariantly constant ($D-2p-1$)-forms coupled to ($2p-1$) CS forms. The
general expression for the sources --charged with respect to the
corresponding gauge algebra-- is presented, focusing on two special cases: $%
0 $-branes and $(D-3)$-branes.

In any dimension, $0$-branes are constructed as topological defects produced
by a surface deficit of ($D-2$)-sphere in AdS space, and they are not
constant curvature spaces for $D>3$. They correspond to dimensionally
continued black holes with negative mass.

On the other hand, in the case of CS (super) gravities, the ($D-3$)-branes
are naked conical singularities (topological defects) obtained by
identification of points with a Killing vector. In $2+1$\ dimensions,
extremal spinning branes of this type are BPS states. Stable ($D-3)$-branes
are shown to exist also in higher dimensions, as well.

Classical field equations are also\ discussed and in the presence of sources
there is a large number of inequivalent and disconnected sectors in solution
space.
\end{abstract}

\maketitle

\section{Introduction}

Chern-Simons (\textbf{CS}) theories are a remarkable family of metric-free,
background-independent, generally covariant gauge theories that extend the
usual concept of minimal coupling between a current and the electromagnetic
potential. Even in the simplest three-dimensional case where CS theory do
not possess local degrees of freedom and is purely topological, it leads to
a classification of three-dimensional manifolds \cite{Witten1}, and gives an
exactly solvable quantum theory of gravity \cite{Witten2}. Quantum CS theory
describes the quantum Hall effect \cite{Zhang-Hansson-Kivelson} and a CS
term provides an alternative gauge-invariant procedure of mass generation
\cite{Deser-Jackiw-Templeton}. If coupled to dynamical matter fields, CS
theories exhibit spontaneous symmetry breaking and a Higgs mechanism that
differs significantly from theories coupled to Yang-Mills fields \cite%
{Higgs-CS}.

There exist Chern-Simons formulations for gravity and supergravity in all
odd dimensions \cite{Review,CS-gravity-3D,CS-gravity} that are truly gauge
theories with fibre bundle structure, which make them good candidates to
tackle the quantization of gravity problem. On the other hand, CS theories
have no adjustable couplings: all the constants in the lagrangian are either
particular combinatorial coefficients, or fixed by quantization. These are
not coupling constants in the standard sense that could be used to define a
perturbative quantum expansion. Moreover, the action is completely
scale-invariant, devoid of dimensionful coefficients or coupling constants
that could get renormalized, and power-counting renormalizable.

In order to define the path integral formulation, it is necessary to couple
the CS connection to external sources, and two natural options present
themselves: \emph{i}) to embed the symmetry as a subalgebra in a larger
gauge algebra, as in CS supergravities, and \emph{ii}) to add a minimal
coupling of the form $\langle jA\rangle $, where $j$\ is covariantly
constant, $Dj=0$. The first option does not solve the problem, it just
changes the setting to a larger gauge group. The second alternative seems
acceptable but does not allow for more general couplings, for instance to
branes, which would be necessary to make contact and compare with the
results in string theory.

Among the different branes, BPS ones are extended objects that define
natural localized vacua because they couple to the fields in a way that
partially respects supersymmetry. This ensures their stability and makes
them acceptable candidate ground states for the perturbative expansion of
the theory expected to describe low energy phenomenology. Since the low
energy limit of the gravitational sector of String theory should contain
higher powers of curvature \cite{String-Lovelock}, here we will consider a
particular case of Lovelock theories known as CS (super-)gravities, which
have the added advantage of being genuine gauge theories with fiber bundle
structure \cite{CS-gravity}.

CS supergravities are interesting systems that offer a natural way to
combine the gravitational field with other forms of matter and interactions
under a unified scheme. However, coupling CS theories to branes in the same
way that one does for standard supergravity as the low energy description of
string theory, proves inconsistent. The problem stems from the clash between
the fields that are expected to be turned on by the tensorial structure of
the branes and the CS dynamics that requires some specific components of the
connection to be present in order to ensure supersymmetry \cite%
{Edelstein-Zanelli}.

In standard supergravity the gravitino\ transforms as a covariant-like
derivative of a spinor, including some curvature components as pieces of the
connection. These extra terms allow nontrivial solutions if a chirality
condition is fulfilled. On the other hand, in the CS theory the gravitino
transforms as a pure covariant derivative of a spinor, which does not allow
a similar solution. The conclusion is then that the naive minimal coupling
between CS supergravity and branes cannot be consistent and still respect
supersymmetry \cite{Edelstein-Zanelli}.

This negative conclusion could be seen as a no-go theorem or simply as one
more indication of the fact that CS theories are exceptions to most of the
standard rules of quantum field theory. In fact, the scope of this article
is to show that a nonstandard coupling does exist between CS theories and
that branes could respect supersymmetry. This claim is based on the
observation that CS theories themselves have a structure that generalizes
the minimal coupling between a gauge connection and a (point) particle to
the case where the particle is replaced by an extended object.

In three-dimensional gravity with negative cosmological constant, it was
shown that topological defects with angular deficit in AdS space
corresponding to static or spinning $0$-branes, represent point-like
external currents that couple in a gauge-invariant way to gravity. These
topological defects are with negative energy states in the BTZ black
hole-like spectrum. Their energy range is between the AdS space ($M=-1$) and
the zero mass black hole, and if spinning, they become stable BPS $0$-branes
in CS supergravity for $M=-|J|$\ \cite{NakedBTZ} (see also \cite{Castro} for
a discussion related to exact solutions and thermodynamics of this kind of
sources).

The presence of non-dynamical external sources leads\ to explicit breaking
of spacetime symmetries, as well as to spontaneous symmetry breaking,\ when
the dynamical matter that couples to the gauge connection has nonzero vacuum
expectation value. Since branes generically violate global (super)AdS
symmetry, and in particular Lorentz and translational invariance, both
bosonic and fermionic Goldstone modes are present in the theory \cite{Tong}.
Furthermore,\ since in the absence of sources the action is gauge invariant,
external matter couplings can lead to a Higgs mechanism,\ and that the
unbroken symmetries in nonabelian case cannot be extended globally due to
winding around a solution.\textbf{\ }Vortices and symmetry breaking in CS
theories with matter\ have been discussed in\textbf{\ }\cite%
{CS-Vortices,Dunne,CSsymmetryBreaking}.

In what follows we focus on possible generalizations of the results in $2+1$
interactions, that is, gauge-invariant couplings between external $p$-brane
sources and CS (super-)gravities, assuming that the sources are static and
non-dynamical.

\section{CS theory as generalized minimal coupling}

Given a connection $A$, a CS action in $2n+1$ dimensions is defined as
\begin{equation}
I_{\text{CS},2n+1}\left[ A\right] =\kappa \int\limits_{M^{2n+1}}\langle
\mathcal{C}_{2n+1}(A)\rangle \,,  \label{2n+1CS}
\end{equation}%
where $M^{2n+1}$ is a ($2n+1$)-dimensional manifold, not necessarily endowed
with a metric structure, the level $\kappa $ is a dimensionless constant and
\begin{equation}
\langle \mathcal{C}_{2n+1}(A)\rangle =\frac{1}{n+1}\,\langle A\left(
dA\right) ^{n}+c_{1}\,A^{3}\left( dA\right) ^{n-1}+\cdots
+c_{n}\,A^{2n+1}\rangle \,.  \label{CSLagrangian}
\end{equation}%
Here\ $A$\ stands for a $1$-form with values in a certain Lie algebra $G$, $%
\langle \cdots \rangle $ denotes an invariant symmetric trace in the Lie
algebra, and $c_{1},\ldots ,c_{n}$ are dimensionless rational coefficients
uniquely determined by the condition that defines a Chern-Simons form,
\begin{equation}
d\langle \mathcal{C}_{2n+1}(A)\rangle =\frac{1}{n+1}\,\langle F^{n+1}\rangle
\,,  \label{CS density}
\end{equation}%
where $F=dA+A^{2}$ is the curvature \cite{Nakahara,Review}. Exterior product
is understood throughout, and the indices will be made explicit when needed.

It is clear from the construction that a CS action has no arbitrary
constants apart from $\kappa $ \cite{Quantized}. This is at the same time a
virtue and a curse: the absence of free parameters means that quantization
cannot renormalize the action (the beta function must vanish), lest gauge
invariance is broken. On the other hand, it becomes extremely difficult to
implement a working perturbative approximation to couple CS actions to
matter sources and it can be seen that the standard strategy that allows to
couple supergravity to branes of various dimensions does not yield the
desired results \cite{Edelstein-Zanelli}.

\subsection{Review: \textbf{Abelian }$p$-\textbf{branes}}

The interaction between gauge fields and matter\textbf{\ }is provided by the
standard minimal coupling recipe,%
\begin{equation}
I^{\text{EM}}=\int\limits_{M^{D}}d^{D}x\,j^{\mu }(x)A_{\mu }(x)\,,
\label{Minimal}
\end{equation}%
where the $j^{\mu }(x)$ is the current produced by a point particle, charged
with respect to the gauge group $U(1)$. The essential feature that selects (%
\ref{Minimal}) among all possible interaction terms is, apart from
simplicity, gauge invariance. If the field $A$\ transforms as a connection, $%
A\rightarrow A+d\Lambda $, $I^{\text{EM}}$\ remains invariant provided $%
j^{\mu }$\ is localized in space and conserved, $\partial _{\mu }j^{\mu }=0$%
. The current density of point charge $j^{\mu }(x)=q\,\dot{z}^{\mu }\delta
^{(2n)}\left( x-z(\tau )\right) $ --where $q$ is the magnitude of the
electric charge and $z^{\mu }(\tau )$ represents its position along the
worldline, parameterized by the affine parameter $\tau $--, satisfies both
requirements.

A point particle can also be viewed as a $0$-brane whose time evolution is a
one-dimensional manifold that supports the current: (\ref{Minimal}) is the
integral of the 1-form $A$\ over the particle's history. The current $j^{\mu
}$\ is also the dual of a ($D-1$)-form $j_{[0]}$\ that projects onto the
worldline of the $0$-brane. For a point source at rest at the origin and
splitting spacetime between the worldline $\Gamma ^{1}$\ and the transverse
space $T^{D-1}$, the source in (\ref{Minimal}) can be replaced by the $(D-1)$%
-form Dirac delta,%
\begin{equation}
j_{[0]}=q_{_{0}}\,\delta (T^{D-1})\,d\Omega ^{D-1}\,,  \label{0brane-source}
\end{equation}%
where $d\Omega ^{D-1}$\ is the volume form in $T^{D-1}$\ (the Dirac delta $%
\delta (x)d^{n}x$\ in $R^{n}$\ is naturally defined as an $n$-form that is
ready to be integrated). Then, (\ref{Minimal}) could also be written as
\begin{equation}
I_{\text{$0$-brane}}[A,j]=\int\limits_{M^{D}}j_{[0]}\wedge
A=q_{0}\int\limits_{\Gamma ^{1}}A=q_{0}\int\limits_{\Gamma ^{1}}A_{\mu
}\,dz^{\mu }\,,  \label{Minimal-0brane}
\end{equation}%
where the conservation law $\partial _{\mu }j^{\mu }=0$, is replaced by the
closure of its dual, $dj_{[0]}=0$. Comparing the second and third
expressions in (\ref{Minimal-0brane}), it is obvious that the current acts
by projecting the integral onto the worldline.

Similar couplings between gauge fields and higher dimensional branes can
also be considered. Attempts to couple gauge fields to branes of different
dimensions were pioneered by Teitelboim \cite{Teitelboim}, where (\ref%
{Minimal}) was generalized assuming extended sources represented by $p$%
-dimensional currents $j^{\mu _{1}\cdots \mu _{p}}$, coupled to $p$-form
fields $A=A_{\mu _{1}\mu _{2}\cdots \mu _{p}}\,dx^{\mu _{1}}\cdots dx^{\mu
_{p}}$. The field strengths (curvature $(p+1)$-form), $F=dA$, are invariant
under Abelian transformations $A\rightarrow A+d\Lambda $, where $\Lambda $
is a $(p-1)$-form. The direct extension of this idea to nonabelian gauge
fields, however, was shown to lead to inconsistencies \cite%
{Teitelboim,Henneaux-Teitelboim}. Our approach circumvents those
difficulties generalizing (\ref{Minimal-0brane}) in a different way: the
minimal coupling can be regarded as a CS action in $0+1$ dimensions and,
analogously, the worldvolume generated by a $2p$-brane can be seen as the
action for a $2p+1$ CS form, Abelian or not \cite{Zanelli}.

The path-dependent but coordinate-independent expression (\ref%
{Minimal-0brane}) is the simplest example of CS action $I_{\text{CS},0+1}[A]$%
,\emph{\ }obtained by setting $n=0$\ in (\ref{CSLagrangian}) in the Abelian
case. Then, an expression analogous to the r.h.s.\ of (\ref{Minimal}) for a
higher dimension can be taken as a $2p+1$\ CS form with support on the
worldvolume generated by the evolution of a $2p$-brane. Thus, the coupling
of an abelian CS connection in $D=2n+1$\ dimensions to a $2p$-brane can be
similarly defined, with the source represented by a $2(n-p)$-form $j_{[2p]}$,%
\begin{equation}
I_{\text{$2p$-brane}}[A,j]=\frac{1}{p+1}\int_{M^{D}}j_{[2p]}\underset{(2p+1)%
\text{-CS form}}{\underbrace{AdA\cdots dA}}\,.  \label{Minimal-2pbrane}
\end{equation}

This is an electromagnetically charged $2p$-brane coupled to a $(2p+1)$
Abelian CS form, where the current $j_{[2p]}$ is the $(2n-2p)$-form,%
\begin{equation}
j_{[2p]}=q_{2p}\,\delta (T^{2n-2p})d\Omega ^{2(n-p)}=q_{2p}\,\delta \left(
x-z\right) \,dx^{1}\wedge \cdots \wedge dx^{2n-2p}\,,
\end{equation}%
with $z$\ labeling the points on the worldvolume of the $2p$-brane, $z\in
\Gamma ^{2p+1}$. The fact that the form $j_{[2p]}$\ is closed implies that
its dual, the current density $j^{\mu _{1}\cdot \cdot \cdot \mu _{2p+1}}$,
is conserved,\ $\partial _{\nu }j^{\nu \mu _{2}\cdot \cdot \cdot \mu
_{2p+1}}=0$.

These sources are easily understood in two extreme cases, namely, $p=0$\ and
$2p=D-3$\ $(p=n-1)$. As discussed above, the first case describes a point
singularity in the spatial section, whose worldvolume is a one-dimensional
line; the second case is a brane whose worldvolume is a manifold of
co-dimension 2 (conical defect). A few explicit examples of classical
solutions for some sources are presented in Appendix \ref{Abelian}.

\subsection{Nonabelian generalization}

The coupling between a nonabelian gauge field $A$ and a nonabelian $(2n-2p)$%
-form source $j_{[2p]}$ that generalizes (\ref{Minimal-2pbrane}) is
\begin{equation}
I_{2p\text{-brane}}[A,j]=\kappa \int\limits_{M^{D}}\langle j_{[2p]}\,%
\mathcal{C}_{2p+1}(A)\rangle \,.  \label{coupling}
\end{equation}%
The $2p$-brane source is given by
\begin{equation}
j_{[2p]}(x)=q_{2p}\,\delta (T^{D-1})d\Omega ^{D-1}\,G^{K_{1}\cdots K_{n-p}}\,%
\text{\ ,}  \label{generic j}
\end{equation}%
where the indices ${\small K}_{1}{\small ,K}_{2}{\small ,\cdots }$ label the
generators of the Lie algebra $\mathcal{G}$, and the operator $%
G^{K_{1}\cdots K_{n-p}}$ is a tensor in the corresponding representation. It
is not guaranteed that the trace $\langle \cdots \rangle $ in (\ref{coupling}%
) yields a nontrivial result; a matching between the Lie algebra, the
invariant trace used and the specific operator $G$ is required to produce
interesting couplings.

The\ natural recipe to couple a gauge connection to a $2p$-brane is: take
the algebra's invariant trace $\langle \cdots \rangle $ and consider any
current of the form (\ref{generic j}). In particular, to couple a CS theory
in $2n+1$ dimensions to a $2p$-brane, we take%
\begin{equation}
I_{2n+1}[A,j]=\kappa \int\limits_{M^{2n+1}}\langle \mathcal{C}%
_{2n+1}(A)-j_{[2p]}\,\mathcal{C}_{2p+1}(A)\rangle \,.  \label{total action}
\end{equation}%
Here we ignore boundary terms, which may be important in order to have a
well-defined finite action principle and conserved charges \cite%
{Miskovic-Olea} and in the quantum theory. The field equations obtained from
the action (\ref{total action}) are
\begin{equation}
F^{n}=j_{[2p]}\,F^{p}.  \label{e.o.m.}
\end{equation}%
Thus, off the worldvolume of the $2p$-brane, the gauge field is a solution
of the source-free field equations $F^{n}=0$, but on the worldvolume other
options exist (see Appendix \ref{Abelian}).

Clearly, any value of $p$ in the range $0\leq p\leq (D-1)/2$ is allowed, and
therefore one is naturally led to consider the most general coupling between
a CS connection and all possible $2p$ branes,%
\begin{eqnarray}
I_{2n+1}[A,j] &=&\kappa
\int\limits_{M^{2n+1}}\sum\limits_{p=0}^{n}(-1)^{n-p}\langle j_{[2p]}\,%
\mathcal{C}_{2p+1}(A)\rangle  \notag \\
&=&\kappa \sum\limits_{p=0}^{n}(-1)^{n-p}\,q_{p}\int\limits_{\Gamma
^{2p+1}}\langle \mathcal{C}_{2p+1}(A)\rangle \,,  \label{general coupling}
\end{eqnarray}%
where the alternating sign $(-1)^{n-p}$\ is introduced to simplify the form
of the field equations. Here we have taken the simplest case of a static
flat brane. A more realistic picture would include dynamically evolving
branes that could also intersect, overlap, and even become embedded into
each other.\

Interaction of branes would require terms in (\ref{general coupling}) that
combine CS densities with different $p$, that in general is not
straightforward to construct as it would involve some generalization of
transgression forms \cite{Moraetal}. Intersecting branes with independent CS
actions living on the different component bubbles were studied in \cite{WZ}.
This sort of \textquotedblleft foam\textquotedblright\ can be viewed as a
formal sum of CS forms integrated over a chain complex, in a system where
the distinction between free theory\ and interaction terms\ is rather
conventional. The fact that the coupling to branes enters on equal footing
with the bulk action suggests a sort of \textquotedblleft
democracy\textquotedblright\ between brane worldvolumes and target space in
CS theories \cite{Mora-Nishino}.

\subsection{Symmetry breaking}

As mentioned above, off the brane the current is covariantly constant, $%
Dj\equiv dj+[A,j]=0$\ (that is, the dual current density is covariantly
conserved, $D_{\nu }j^{\nu \mu _{2}\cdot \cdot \cdot \mu _{2p+1}}=0$), which
ensures gauge invariance of (\ref{coupling}) from the point of view of an
external observer living in $M^{D}$. On the brane the gauge invariance would
be reduced to the subalgebra $\mathcal{\tilde{G}}\subset \mathcal{G}$\
spanned by those generators that commute with $G^{K_{1}\cdots K_{n-p}}$\ ,
but this is not an issue for the dynamics off the brane's worldvolume.

On the other hand, the presence of the non-dynamical source $j_{[2p]}$\ as a
fixed feature in the ambient space $M^{D}$\ does reduce the spacetime
symmetries to those transformations that leave the source unchanged. This
symmetry can be restored if the current is produced by some other particles
or fields whose dynamics is included in the same action principle.

It is worthwhile noticing that the presence of dynamic matter coupled to a
CS connection leads to spontaneous symmetry breaking when its vacuum
expectation value is nonvanishing. If the broken symmetry is global, as for
the brane that is non-invariant under spatial translations and Lorentz
rotations, then massless Goldstone modes are present in the theory. In a
supersymmetric extension of the theory, fermions may also contribute to the
zero modes (see Ref. \cite{Tong}).

On the other hand, since CS is a gauge theory, nonabelian coupling to
external matter may result in symmetry breaking of Schwarz type where the
unbroken symmetries cannot be extended globally \cite{Schwarz}, or in a
Higgs mechanism. Vortices are a class of string-like solutions\ carrying
magnetic flux confined in their center that arise in couplings, e.g., with
scalar fields \cite{Vortex}, and whose existence may make unbroken
symmetries multivalued around the string; these are Alice strings \cite%
{Schwarz,Alford-Benson-Coleman-MarchRussell} possessing non-localized
electric charge (Cheshire charge), thus related to $p(>0)$-branes. Vortices
in CS theory are discussed, for example, in \cite{Dunne,CSsymmetryBreaking}
and further analysis of Higgs mechanism in CS theories is analyzed in Refs.
\cite{CS-Vortices}. The issue of symmetry breaking is an open problem to be
discussed in depth elsewhere.

\section{Coupling branes to CS gravity}

CS gravities are theories where the gauge symmetry is the invariance group
of the local tangents to the spacetime manifold; their supersymmetric
extensions are CS supergravities \cite{CS-gravity}. The simplest of such
theories occurs for $D=2+1$, where it was observed that 0-branes
corresponding to topological defects are naked singularities in the negative
energy spectrum of the BTZ black hole \cite{Miskovic-Zanelli}. It turns out
that the extremal spinning 0-branes of this sort are BPS states. Here we
analyze the generalization of this picture to higher dimensions.

\subsection{CS gravity}

The cases in which the algebra $\mathcal{G}$ is either $so(D,1)$, $so(D-1,2)$
or $iso(D-1,1)$ represent an important class of CS theories that describe
the dynamics of spacetime (gravitation) with positive, negative or vanishing
cosmological constant, respectively \cite{Review}. If the trace $\langle
\cdots \rangle $ is given by the Levi-Civita tensor, these gravitation
theories are particular cases of Lovelock theories, which take the form \cite%
{Lovelock}%
\begin{equation}
L=\sum\limits_{p=o}^{[D/2]}\alpha _{p}\,\epsilon _{a_{1}a_{2}\cdots
a_{D}}\,R^{a_{1}a_{2}}R^{a_{3}a_{4}}\cdots
R^{a_{2p-1}a_{2p}}e^{a_{2p+1}}\cdots e^{a_{D}}\,.  \label{Lovelock}
\end{equation}

This Lagrangian describes the most general $D$-dimensional gravitation
theory for the metric, if the spacetime is assumed to be Riemannian
(torsion-free). For spacetimes of $D=2n+1$ dimensions, the Lagrangian takes
the CS form if the coefficients $\alpha _{p}$\ are chosen as $\alpha _{p}=%
\frac{\left( \mp \ell ^{-2}\right) ^{p}}{D-2p}\binom{n}{p}$, where $\ell $\
is the (A)dS radius. The cosmological constant is $\Lambda =\pm \ell ^{-2}$\
for the de Sitter ($+$), the anti-de Sitter ($-$), and $\Lambda =0$\ for the
Poincar\'{e} (if $\alpha _{p}=\delta _{p}^{n}$) algebras, respectively (see,
e.g., \cite{Review,CS-gravity-3D,CS-gravity}).

The general Lovelock Lagrangians are gauge theories for the Lorentz group $%
SO(D-1,1)$\ and, in contrast with CS theories, they are not gauge theories
for $SO(D-1,2)$, $SO(D,1)$, or $ISO(D-1,1)$, since the fields in the action
are not the components of the connection for the respective algebras.
Besides the connection $\omega _{\text{ \ }b}^{a}$, Lovelock actions contain
the vielbein $e^{a}$, which transforms in the vector representation of the $%
SO(D-1,1)$. The vielbein cannot decouple from the gauge connection, so
Lovelock theories in general --and General Relativity in particular--, are
nonabelian systems like QCD with $e^{a}$\ playing the role of quarks, but
whose matter-free limit does not exist. Moreover, in the generic Lovelock
theories the dimensionful parameters $\alpha _{p}$ can take arbitrary values
and can get renormalized in the quantum theory because they are not
protected by gauge invariance.

The functions $R_{\text{ \ }b}^{a}=d\omega _{\text{ \ }b}^{a}+\omega _{\text{
\ }c}^{a}\,\omega _{\text{ \ }b}^{c}$ , $e^{a}$ and $\epsilon
_{a_{1}a_{2}\cdots a_{D}}$ are Lorentz tensors, which makes all Lovelock
theories --including Einstein gravity-- invariant under local Lorentz
transformations. CS theories enjoy an enhanced gauge symmetry that results
from particular choice of $\alpha _{p}$. By virtue of this choice, $e^{a}$
and $\omega _{\text{ \ }b}^{a}$\ can be combined into a single connection
for the corresponding gauge group, all dimensionful constants can be
absorbed in the fields and the result is a fully scale-invariant
gravitational action, where all the coupling constants are fixed rational
numbers.

We shall mostly focus on Chern-Simons-AdS gravity, described by the action (%
\ref{total action}) for the AdS algebra $so(D-1,2)$, where the torsion
tensor may enter the lagrangian explicitly. The AdS generators are
antisymmetric matrices $J_{AB}$ ($A,B=0,\ldots ,D=2n+1$) acting on vectors
of the local tangent to the spacetime manifold, an abstract covering space
with the metric $\eta _{AB}=(-,+,\ldots ,+,-)$. In this representation, the
Lie algebra $so(D-1,2)$\ reads%
\begin{equation}
\lbrack J_{AB},J_{CD}]=\eta _{AD}J_{BC}-\eta _{AC}J_{BD}-\eta
_{BD}J_{AC}+\eta _{BC}J_{AD}\,.  \label{Lie-algebra}
\end{equation}%
The symmetrized trace of the product of $n+1$ of these generators is given
by the Levi-Civita invariant tensor,%
\begin{equation}
\langle J_{A_{1}B_{1}}\cdots J_{A_{n+1}B_{n+1}}\rangle =\epsilon
_{A_{1}B_{1}\cdots A_{n+1}B_{n+1}}\,.
\end{equation}%
Using the decomposition of AdS indices $A=(a,2n+1)$, $a=0,1,...,2n$, and
calling $J_{a}:=J_{a\,2n+1}$, the gauge field reads%
\begin{equation}
A=\frac{1}{2}\,\omega ^{ab}J_{ab}+\frac{1}{\ell }\,e^{a}J_{a}\,,  \label{A}
\end{equation}%
and the corresponding AdS curvature is
\begin{equation}
F=\frac{1}{2}\,\left( R^{ab}+\frac{1}{\ell ^{2}}\,e^{a}e^{b}\right) J_{ab}+%
\frac{1}{\ell }\,T^{a}J_{a}\,,  \label{F}
\end{equation}%
where $T^{a}\equiv De^{a}=de^{a}+\omega _{\text{ \ }b}^{a}\,e^{b}$ is the
torsion $2$-form. In case of CS supergravity, the gauge connection (\ref{A})
contains the additional components along the supersymmetric extension of the
AdS algebra, that contain fermionic generators, and may include also bosonic
ones as required by the closure the superalgebra \cite{Super}.

\subsection{Gravitational sources}

In the Abelian case, there are very few requirements for $j$ to define an
acceptable current source: it must have support on a $2p$-brane that
generates a $(2p+1)$-dimensional submanifold of spacetime, and to be
conserved in order to respect gauge invariance. In the nonabelian case,
since the charge is not\textbf{\ }invariant, but transforms irreducibly
under the gauge group, the gauge invariance of the interacting theory may
have fewer local symmetries than the source-free CS action.

Let us consider a static $0$-brane sitting at the origin in a global AdS$%
_{2n+1}$ spacetime. The invariance group of the brane worldline $\Gamma ^{1}$
includes time translations, $SO(1,1)$, and spatial rotations, $SO(2n)$. The
only operator in the AdS algebra $so(2n,2)$ that commutes with time
translations --or rather AdS boosts in the time direction--, and spatial
rotations is
\begin{eqnarray}
G &=&\frac{1}{\left( 2n\right) !}\,\epsilon ^{0A_{1}B_{1}\cdots
A_{n}B_{n}2n+1}\,J_{A_{1}B_{1}}\cdots J_{A_{n}B_{n}}\,
\label{0-brane source} \\
&=&\frac{1}{\left( 2n\right) !}\,\epsilon ^{\alpha _{1}\beta _{1}\cdots
\alpha _{n}\beta _{n}}\,J_{\alpha _{1}\beta _{1}}\cdots J_{\alpha _{n}\beta
_{n}}\text{\ },
\end{eqnarray}%
where the tangent space indices $\alpha $,$\beta $,\ correspond to the
tangent space of the spatial section of the manifold, so that $A=\left\{
0,\alpha ,2n+1\right\} $. For example, in three dimensions, the $SO(2)$
generator is just the rotation $G=J^{12}$ in the 1-2 plane, whereas in five
dimensions this is a spherically symmetric combination $G=\frac{1}{3}%
\,\left( J^{12}J^{34}-J^{13}J^{24}+J^{14}J^{23}\right) =\frac{1}{24}%
\,\epsilon _{\alpha \beta \gamma \rho }\,J^{\alpha \beta }J^{\gamma \rho }$.
The fact that $G$ commutes with all generators of spatial rotations, $\left[
G,J_{\alpha \beta }\right] =0$, follows directly from the identity
\begin{equation}
J_{\beta }^{\alpha _{1}}\epsilon ^{\beta \alpha _{2}\cdots \alpha
_{2n}}+\cdots +J_{\beta }^{\alpha _{p}}\epsilon ^{\alpha _{1}\cdots \alpha
_{p-1}\beta \alpha _{p+1}\cdots \alpha _{2n}}+\cdots +J_{\beta }^{\alpha
_{2n}}\epsilon ^{\alpha _{1}\cdots \alpha _{2n-1}\beta }=0\text{\thinspace\ .%
}
\end{equation}

From (\ref{0-brane source}), the symmetric invariant trace defining the
interaction of the AdS connection and the $0$-brane is
\begin{equation}
\left\langle GJ_{AB}\right\rangle =\delta _{\lbrack AB]}^{[0\text{\ }%
2n+1]}\,.
\end{equation}

In this way, the introduction of a $0$-brane breaks the symmetry from $%
SO(2n,2)$ down to $SO(1,1)\times SO(2n)$, with a net loss of $\Delta =4n$
symmetry generators.

In the case of a $2p$-brane in AdS$_{2n+1}$, a worldvolume is a $(2p+1)$%
-dimensional time-like manifold $\Gamma ^{2p+1}$ that could have at most $%
SO(2p,2)$ symmetry on its tangent space. The $(2n-2p)$-dimensional spacelike
transverse section can have a tangent space with local $SO(2n-2p)$
invariance. In this case the reduction is from $SO(2n,2)$ down to $%
SO(2p,2)\times SO(2n-2p)$, with a loss of $\Delta =4(n-p)(p+1)$ symmetry
generators. The largest proper $2p$-brane that can be coupled this way is
one with $p=n-1$, in which case the $\Delta =4n$,\ as for the case with $p=0$%
.

This analysis also applies to cases where the gauge group for the CS action
without external sources is dS, Poincar\'{e}, or a supersymmetric\ extension
of Poincar\'{e}, but this point will not be analyzed further here.

\subsection{$0$-branes in $2n+1$ dimensions}

The 2+1 CS theory of gravity for the AdS group $SO(2,2)$\ is the simplest
analog of General Relativity that is realistic enough to capture some of its
essential features. Although this gravitational toy model has no Newtonian
attraction, it gives rise to nontrivial black hole solutions that in many
ways resemble astronomic black\ holes at many galactic nuclei \cite{BTZ}.
For a comprehensive review, see \cite{Carlip}.

In three dimensions, the only $2p$-brane that can be properly embedded is a
point source in the two-dimensional spatial section, a $0$-brane. The
resulting spacetime is a negative energy naked singularity, generated by a
topological defect in spacetime, similar to a string of angular deficit in a
three-dimensional crystal \cite{NakedBTZ}\textbf{.} For $D=2n+1>3$, gravity
with negative cosmological constant\ is described by a CS action for the $%
so(D-1,2)$ algebra, with connection (\ref{A}). In this case, a point-like
source is a $0$-brane\emph{\ }describing a spherically symmetric topological
defect, produced by a surface deficit on a $(D-2)$-dimensional sphere.\ This
geometry is given by the classical solution
\begin{equation}
ds^{2}=-\left( 1+\frac{r^{2}}{\ell ^{2}}\right) \,dt^{2}+\frac{dr^{2}}{1+%
\frac{r^{2}}{\ell ^{2}}}+\left( 1-\alpha \right) ^{2}r^{2}\,d\Sigma
_{D-2}^{2}\,,  \label{0-brane metric}
\end{equation}%
where $d\Sigma _{D-2}^{2}$ is a metric of the unit\ $(D-2)$-sphere. For $%
\alpha =0$\ the metric reduces to that of the global AdS geometry, whereas
for $0<\alpha <1$\ there is a defect of magnitude $\alpha \Sigma (S^{D-2})$,
where $\Sigma (S^{D-2})$\ is the surface area of the whole sphere, and the
metric exhibits a conical singularity at $r=0$. This metric includes the $%
2+1 $ case where the topological defect is generated by an identification
along a Killing vector in an Euclidean $x^{1}$-$x^{2}$-plane with a fixed
point at $r=0$ , producing a conical singularity in this plane. For $D>3$,
the topological defect is not produced by a Killing vector identification
and it changes the local geometry of spacetime and the curvature is not
constant for $\alpha \neq 0$. It is straightforward to show that, for $r\neq
0$, the geometry has non-vanishing AdS curvature,%
\begin{equation}
R^{pq}+\frac{1}{\ell ^{2}}\,e^{p}e^{q}=-\frac{\alpha (\alpha -2)}{%
r^{2}\left( 1-\alpha \right) ^{2}}\,e^{p}e^{q}\,\text{ \ \ \ \ \ for }r\neq 0%
\text{ },  \label{Riemann}
\end{equation}%
where the tangent space indices $p,q,\ldots $ correspond to the angular
directions, so that the right hand side in (\ref{Riemann}) vanishes for $%
D\leq 3$. The curvature is negative and not constant for $D>3$ and, for $r>0$%
, the Ricci scalar is
\begin{equation}
R=-\frac{D\left( D-1\right) }{\ell ^{2}}-\frac{(D-2)\left( D-3\right) \alpha
(\alpha -2)}{r^{2}\left( 1-\alpha \right) ^{2}}\,.
\end{equation}%
Changing coordinates as $\left( r,t\right) =\left( \frac{\rho }{1-\alpha },%
\text{ }(1-\alpha )\tau \right) ,$ the metric (\ref{0-brane metric}) becomes%
\begin{equation}
ds^{2}=-\left( (1-\alpha )^{2}+\frac{\rho ^{2}}{\ell ^{2}}\right) d\tau ^{2}+%
\frac{d\rho ^{2}}{(1-\alpha )^{2}+\frac{\rho ^{2}}{\ell ^{2}}}+\rho
^{2}\,d\Sigma _{D-2}^{2}\,.  \label{D-dim BH}
\end{equation}%
For $D=3$, this metric has the form of the 2+1 black hole, but with negative
mass,\ a naked singularity produced by a static $0$-brane \cite{NakedBTZ}.
For $D=2n+1\geq 5$, this solution describes dimensionally continued
\textquotedblleft black holes with negative mass\textquotedblright\ ($-1<M<0$%
) \cite{DimContinued}. The mass is related to the magnitude of the defect as%
\begin{equation}
(1-\alpha )^{2}=1-\left( 1+M\right) ^{\frac{1}{n}}\,.
\end{equation}

From (\ref{Riemann}) and (\ref{D-dim BH}) it is clear that for $\alpha =0$\ (%
$M=-1$), the AdS space geometry is recovered. (For $\alpha =2$\ the geometry
again has constant curvature, since the metric (\ref{0-brane metric})
possesses a discrete symmetry $\left( 1-\alpha \right) \rightarrow -\left(
1-\alpha \right) $, so that both $\alpha =0$ and $\alpha =2$ have no
defect.) The surface deficit corresponds to $\alpha $ in the range $\alpha
\in (0,1)\cup (1,2)$.

For $\alpha \neq 0$, the naked singularity is a static spherically symmetric
configuration with mass in the range $-1<M<0$. For $M<-1$, the geometric
interpretation\ becomes obscure as it would represent an angular sector
greater than that of a full solid angle. It is unclear whether naked
singularities of this type could exist at all.

In order to find the source generating this geometry, the AdS curvature for
the metric (\ref{0-brane metric}) must be computed, regularizing it at $r=0$%
. If the regulated curvature is $F_{\epsilon }$, the source is given by $%
j=\lim_{\epsilon \rightarrow 0}F_{\epsilon }^{n}$. The result is (see
Appendix \ref{0 brane regularization})
\begin{equation}
\left\langle j(x)J_{AB}\right\rangle =q_{0}^{(D)}\,\delta ^{(D-1)}\left(
\vec{x}\right) \,dx^{1}\cdots dx^{D-1}\,\delta _{\lbrack AB]}^{[0D]}\,,
\label{0-source}
\end{equation}%
where the \textquotedblleft charge\textquotedblright\ $q_{0}^{(D)}$ in $D$
dimensions is a polynomial of order $D-2$ in the constant $\alpha $, given
by
\begin{eqnarray}
q_{0}^{(3)} &=&2\pi \alpha \,, \\
q_{0}^{(5)} &=&\frac{4\pi }{3}\,\alpha ^{2}\,\left( 3-\alpha \right) \,,%
\text{etc.}
\end{eqnarray}%
For the general form of the charge see Appendix \ref{0 brane regularization}%
. Similar spherically symmetric angular defects can also be introduced in de
Sitter (dS) and in flat space, leading to geometries similar to (\ref{D-dim
BH}), but with the metric functions $f_{\text{dS}}^{2}=\rho ^{2}/\ell
^{2}-(1-\alpha )^{2}$ and $f_{\text{flat}}^{2}=\rho ^{2}/\ell ^{2}$,
respectively. The dS case in $2+1$ dimensions was discussed in \cite%
{Reznik92}.

The static $0$-branes described above admit no globally defined, covariantly
constant spinors and are therefore not BPS states --in absence of other
fields that could couple to spinors--, except in the trivial case $M=0$. In $%
3D$, however, the source (\ref{0-source}) of the form $j=2\pi \alpha
\,\delta ^{(2)}\left( T_{12}\right) \,dx^{1}dx^{2}\,J_{12}\,$(here $x^{A}$, $%
A=0,\ldots ,3$, are the coordinates in the embedding flat space $\mathbb{R}%
^{2,2}$ and $J_{12}$ is the AdS$_{3}$ generator), can be extended to the
full Cartan subalgebra of AdS$_{3}$ generated by $J_{03}$ and $J_{12}$,
allowing for the existence of two conserved charges, related to the mass and
angular momentum $M,J$ of the 0-brane. In the extreme case,\ $\left\vert
M\right\vert \ell =\left\vert J\right\vert $, the current
\begin{equation}
j_{\text{ext}}=2\pi \alpha \,\delta ^{(2)}\left( T_{12}\right)
\,dx^{1}dx^{2}\,\left( J_{03}-J_{12}\right) \,,
\end{equation}%
leads to an extreme $0$-brane produced by an identification with a Killing
vector with fixed points at $r=0$. As shown in \cite{NakedBTZ}, the extreme
solution for this brane is a BPS solution admitting one globally defined
Killing spinor, preserving $1/4$ of supersymmetries of AdS, and which
behaves asymptotically as the Killing spinor for zero-mass BTZ black hole
\cite{Coussaert-Henneaux}.

\subsection{Codimension $2$ branes}

In dimensions $D>3$, it is possible to construct higher dimensional $2p$%
-branes, for example introducing an angular deficit in $S^{1}$\ only, that
leads to a geometry describing a (spinning and non-spinning) codimension $2$%
\ brane. The question arises whether those solution would be stable or not.
In two examples in five dimensions, we show that stable BPS 2-branes exist.

\subsubsection{Super CS in AdS space}

Stable BPS solutions in five-dimensional CS-AdS supergravity were found in
Ref.\ \cite{Miskovic-Troncoso-ZanelliBPS} for a supersymmetric extension of
AdS$_{5}$\ algebra $su(2,2|4)$. We show now that a CS gauge connection for
this solution couples to a current of the type (\ref{generic j})\textbf{\ }%
that corresponds to a 2-brane. In this case, the field equations have the
form $FF=jF$.

As shown in \cite{Miskovic-Troncoso-ZanelliBPS}, the purely gravitational
part of the solution is locally AdS (except at singularities) whose spatial
boundary has topology isomorphic to the torus $T^{3}$. The bosonic sector of
CS matter required by supersymmetry is given by $u(1)\times su(4)$
connection. Explicitly, the solution has the form%
\begin{equation*}
A=A_{\text{AdS}}+bG_{1}+a^{12}G_{12}+a^{34}G_{34}\,,
\end{equation*}%
where $A_{\text{AdS}}$\ is the AdS connection,\textbf{\ }and $b$\ and$\
a^{IJ}$,\ are $u(1)$\ and $su(4)$\ gauge fields, respectively. The Abelian
field $b$\ describes currents that generate constant electric and magnetic
fields whose strength has non-vanishing determinant. The generators $G_{12}$%
\ and $G_{34}$\ of $SU(4)$\ commute, therefore the nonabelian gauge field $%
a^{IJ}$\ breaks the symmetry to $U(1)\times U(1)$\ and describes a soliton
that winds up around one handle of the torus $S^{1}$\ at spatial infinity.
Its topological charge is associated to the arbitrary phase $d\theta
=a^{12}-a^{34}$, that carries nontrivial instanton number. The wrapping up
around $S^{1}$\ is obtained by the Killing vector identification and, just
as in 3D case, produces a $\delta $-function\ in the transverse plane that
has support in the 3-dimensional world volume of a 2-brane.

If $dd\theta \sim \delta (T^{2})$\ (nontrivial winding), then $F\sim \delta
(T^{2})$, and consequently $j\sim \delta (T^{2})$; thus the current contains
a combination of the generators $G_{1},$\ $G_{12}$\ and $G_{34}$. It is less
obvious whether there exist components along some other AdS generators $%
J_{AB}$\ as well, since the phase $\theta $\ is\ a general function of the
local coordinates.

The asymptotic symmetries of the super CS-AdS$_{5}$\ theory are described by
the supersymmetric extension of WZW$_{4}$\ algebra\ and this particular
solution corresponds to the ground state saturating the Bogomol'nyi bound.

\subsubsection{2-brane in AdS}

It is possible to construct a solution of the CS equations with an angular
defect in higher dimensions, in analogy with the case in 2+1 dimensions by
making an identification of one angular coordinate $\phi $. In this way, a
brane with a worldvolume of codimension $2$\ is obtained. For simplicity, we
focus again on the five-dimensional case. The AdS$_{5}$ space is given by
the constraint $\,x\cdot x=-\ell ^{2}$ in the embedding six-dimensional flat
space given by the metric $ds^{2}=\eta _{AB}\,dx^{A}dx^{B}$ with Lorentzian
signature$\,\left( -++++-\right) $. Parameterizing the Euclidean planes $%
(x^{0}$-$x^{5})$, $(x^{1}$-$x^{2})$, and $(x^{3}$-$x^{4})$ in a form that
explicitly reproduces the AdS$_{5}$ constraint, similarly to the
three-dimensional expressions in Ref. \cite{NakedBTZ},%
\begin{equation}
\begin{array}{lll}
x^{0}=A\,\cos \phi _{05}\,,\qquad  & x^{1}=B\sin \theta \cos \phi
_{12}\,,\qquad  & x^{3}=B\cos \theta \cos \lambda \,, \\
x^{5}=A\,\sin \phi _{05}\,, & x^{2}=B\sin \theta \sin \phi _{12}\,, &
x^{4}=B\cos \theta \sin \lambda \,,%
\end{array}
\label{x0-x5}
\end{equation}%
where $A^{2}-B^{2}=\ell ^{2}$. Note that this last condition implies that
the origin of the 0-5 plane in the embedding space, that is, $A=0$, is not
part of the AdS spacetime. Here\ $A$ and $B$ are chosen as the following
real functions of the radial coordinate,
\begin{equation}
A=\sqrt{\frac{\rho ^{2}+\ell ^{2}a^{2}}{a^{2}-b^{2}}}\,,\qquad B=\sqrt{\frac{%
\rho ^{2}+\ell ^{2}b^{2}}{a^{2}-b^{2}}}\,,  \label{A,B}
\end{equation}%
where $a$ and $b$ are allowed to take complex values but $a^{2}\neq b^{2}$
(nonextremal case). In this parametrization, the metric takes the form%
\begin{equation}
ds^{2}=-dA^{2}+dB^{2}-A^{2}d\phi _{05}^{2}+B^{2}\left( d\theta ^{2}+\sin
^{2}\theta \,d\phi _{12}^{2}+\cos ^{2}\theta \,d\lambda ^{2}\right) \,.
\label{5D metric}
\end{equation}%
The lapse function can be read off \ directly from the radial part, $%
f^{2}(\rho )=a^{2}+b^{2}+\frac{\rho ^{2}}{\ell ^{2}}+a^{2}b^{2}\,\frac{\ell
^{2}}{\rho ^{2}}$, where $a^{2}+b^{2}$\ can be recognized as the mass
parameter\textbf{,}%
\begin{equation*}
a^{2}+b^{2}=-M\text{ .}
\end{equation*}%
For real $a$\ and $b$, $M$\ is negative (naked singularity) or zero, and the
spin is\ given by\ $J/\ell =2ab$. These are the same formal relations
between ($a$, $b$) and ($M$, $J$) as for the $0$-brane in $2+1$ dimensions.\
In order to compare with more standard forms, one can redefine and $\phi
_{05}$, $\phi _{12}$ as helicoidal coordinates given by%
\begin{equation}
\phi _{05}=b\phi +\dfrac{a\tau }{\ell }\,\text{,\qquad and}\qquad \phi
_{12}=a\phi +\dfrac{b\tau }{\ell }\,\text{.}
\end{equation}

The form of the metric in Schwarzschild-like coordinates $(\tau ,\rho
,\theta ,\lambda ,\phi )$ is complicated and not very enlightening to write
it explicitly here. However, it can be shown that the curvature is constant
for $\rho \neq 0$, and hence, the spacetime is locally AdS, reflecting the
fact that the geometry is AdS$_{5}$\ with an appropriate identification, to
wit, $\phi \simeq \phi +2\pi $.

The identification connects points of the embedding AdS spacetime separated
by the Killing vector $\xi =-2\pi \alpha \,J_{12}+2\pi \beta \,J_{05}$,
where $J_{AB}=x_{A}\,\partial _{B}-x_{B}\,\partial _{A}$ are the AdS
generators. The coefficients $\alpha $\ and $\beta $\ correspond to angular
deficits $\alpha =1-a$\ and $\beta =b$\ in the planes (1-2) and (0-5)
respectively, related to the mass and spin of the solution.

Choosing the vielbein as $e^{0}=A\,d\phi _{05}$, $e^{1}=C\,d\rho $, $%
e^{2}=B\sin \theta \,d\phi _{12}$, $e^{3}=B\,d\theta $\ and\ $e^{4}=B\,\cos
\theta \,d\lambda $, and assuming the spin-connection to be torsionless for $%
\rho \neq 0$, it is straightforward to calculate the\ AdS curvature%
\begin{eqnarray}
F &=&\left( \frac{1}{\ell }\,A\,J_{05}+\dfrac{A^{\prime }}{\sqrt{B^{\prime
2}-A^{\prime 2}}}\,J_{01}\right) \,dd\phi _{05}  \notag \\
&&+\left( -\dfrac{B^{\prime }}{\sqrt{B^{\prime 2}-A^{\prime 2}}}\,\sin
\theta \,J_{12}+\cos \theta \,J_{23}+\frac{1}{\ell }\,B\sin \theta
\,J_{25}\right) \,dd\phi _{12}\,,
\end{eqnarray}%
where the prime denotes radial derivative. Using the identities \cite%
{Jackiw90} $dd\phi _{05}=2\pi \beta \,\delta (T_{05})\,dx^{0}dx^{5}$, $%
dd\phi _{12}=$ $-2\pi \alpha \,\delta (T_{12})\,dx^{1}dx^{2}$, and the field
equations $F(F-j)=0$, we find that there is a sector of CS gravity where the
current is $j=F$. In this sector, the current is
\begin{equation}
j=2\pi b\,G_{05}\,\delta (T_{05})\,dx^{0}dx^{5}+2\pi \alpha \,G_{12}\,\delta
(T_{12})\,dx^{1}dx^{2}\,.  \label{current2-brane}
\end{equation}%
It can be checked that the generators $G_{05}$ and $G_{12}$ are mutually
commuting,%
\begin{eqnarray}
G_{05} &=&\frac{1}{\sqrt{a^{2}-b^{2}}}\,\left( a\,J_{05}+b\,J_{01}\right) \,,
\\
G_{12} &=&\frac{\sin \theta }{\sqrt{a^{2}-b^{2}}}\left(
a\,J_{12}-b\,J_{25}\right) -\cos \theta \,J_{23}\,,
\end{eqnarray}%
which is similar to the case in 2+1 dimensions. Note however that since $%
A\neq 0$, the first term in the current (\ref{current2-brane}) vanishes
identically. Thus, $j$\ is not composed by two 4-dimensional planar sources
(in the embedding space) intersecting on the 3-4 plane, as one could expect
from (\ref{current2-brane}). For the static solution ($b=0$), one obtains%
\begin{equation}
j_{\text{static}}=2\pi \alpha \,\left( \sin \theta \,J_{12}-\cos \theta
\,J_{23}\,\right) \,\delta (T_{12})\,d\Omega _{12}^{2}\,.
\end{equation}%
Clearly, both in the static and spinning cases there is a conical
singularity at$\ \rho =0$, and also like in the 2+1 case, the extremal
2-brane can be constructed as the limit $a=b$\ ($A/B=1$\ or $\left\vert
M\right\vert \ell =\left\vert J\right\vert $).

In general, for codimension 2 branes, the field equations are $%
F^{n}=j_{[2n-2]}\,F^{n-1}$, where the current $j_{[2n-2]}$ is a 2-form, and
there always exist a sector where $j_{[2n-2]}=F$, and $F$ corresponds to a
conical singularity in a two-dimensional plane. Ideas similar in spirit were
recently discussed in \cite{Cuadros-Melgar et al}.

The nonextremal massive spinning 2-branes in five dimensions need not be BPS
states. However, one might expect that stable (BPS) configurations can be
constructed as extremal spinning 2-branes\ in analogy with the
three-dimensional case. For example, a restriction of the metric (\ref{5D
metric}) to a submanifold $\theta =\pi /2$, $\lambda =1$\ is the three
dimensional metric $ds^{2}=-dA^{2}+dB^{2}-A^{2}d\phi _{05}^{2}$\ that
describes a $0$-brane naked singularity in Schwarzschild-like coordinates $%
(\tau ,\rho ,\theta =\frac{\pi }{2},\lambda =1,\phi )$\ \cite{NakedBTZ}. The
extremal spinning naked singularity $a=b$\ ($\left\vert M\right\vert \ell
=\left\vert J\right\vert $) can be similarly embedded in this submanifold,
and as shown in \cite{EGMZ}, this defines a BPS state as well. Moreover,
based on the fact that there exist BPS states in 5-dimensional CS
supergravity \cite{Miskovic-Troncoso-ZanelliBPS} where the space is also
locally AdS, and that the asymptotic isometries of those states are $%
S^{1}\times S^{1}\times S^{1}$, or $S^{3}$, which correspond to the
isometries of static branes, one can conjecture that those BPS states and
the naked spinning codimension two brane are in fact related. However, the
presence of other gauge fields in that theory makes that identification
difficult. Moreover,\textbf{\ }five-dimensional CS supergravity has a very
rich dynamical structure with various disconnected sectors in phase space
characterized by different local symmetries and degrees of freedom \cite%
{Miskovic-Troncoso-Zanelli}. In some of those sectors the AdS space is not a
stable configuration. It would be interesting to see how the extremal BPS
2-brane fits in this scenario, since it might not carry maximal number of
local degrees of freedom.

\section{Summary and prospects}

The minimal coupling between an electric point charge and an electromagnetic
potential is the simplest CS system --the $0+1$ case--, and at the same time
the prototype of how a brane couples to a connection \cite{Zanelli}. Then,
it seems natural to regard any CS system as a form of coupling between a
brane and a (non-) abelian connection. Here we have explicitly shown how
this idea can be exploited to couple $0$- and\ ($2n-2$)-branes to gravity,
when the spacetime geometry is described by a CS action in $D=2n+1$\
dimensions. The cases of other $2p$-branes are technically more complicated,
but in principle can be worked out in a manner similar to these two extreme
examples. Moreover, CS theories can be viewed as describing the dynamics of
a (nonabelian) connection living on the worldvolume of a $2p$-brane, which
is a topological defect in the embedding space.

These topological defects are natural sources for gravity, which in the
particular case of a $0$-brane, has been shown to correspond to the negative
energy spectrum of black holes in gravitational CS theories. This part of
the spectrum corresponds to the gap between the massless black hole\ ($M=0$)
and anti-de Sitter spacetime ($M=-1$).

The fact that these topological defects are naked singularities does not
mean they are necessarily\ unphysical. Moreover, it has been shown that if
endowed with the right amount of angular momentum, they can be stable BPS
objects \cite{NakedBTZ}.\emph{\ }The fact that these are negative energy
states is not contradictory with their supersymmetric nature, because these
are supersymmetric extensions of the AdS --and not the Poincar\'{e}-- group.
Furthermore, it has been argued that negative energy degrees of freedom are
necessary for consistent microscopic description of the entropy of the BTZ
black hole \cite{Krasnov-Carlip}.

The coupling between a connection and a $2p$-brane of the form (\ref%
{coupling}) can exist in any gauge system, described by a CS, or a
Yang-Mills action, or even for some more exotic form of gauge-invariant
action, such as the Born-Infeld theory. An interesting system to study, for
example, could be that of the Maxwell field in 3+1 dimensions coupled to a
2-brane,%
\begin{eqnarray}
I &=&\frac{1}{4}\int F^{\mu \nu }F_{\mu \nu }\,d^{4}x+\int j\wedge A\wedge
dA,  \notag \\
&=&\int \left( \frac{1}{4}\,F^{\mu \nu }F_{\mu \nu }+q\,\delta \left( \Sigma
\right) n^{\alpha }\epsilon _{\alpha \beta \mu \nu }\,A^{\beta }\partial
^{\mu }A^{\nu }\right) d^{4}x\text{ },
\end{eqnarray}%
where $q$\ is the charge and $\Sigma \,$\ is the worldsheet of the membrane.
Obviously, since the worldsheet singles out a direction, $n^{\alpha }$, the
membrane breaks Lorentz invariance, but the $U(1)$\ gauge invariance is
unaffected. This model has been recently proposed in relation to a possible
mechanism for breaking Lorentz invariance \cite{Jackiw05}.

The interaction term (\ref{coupling}) is interesting in CS gravity because
it is a natural gauge-invariant coupling which does not require to introduce
a metric. Moreover, CS gravity is a gauge theory where gauge invariance
reflects the symmetries of the local tangents to spacetime. Hence, the
presence of the brane reduces the AdS symmetry of the $D$-dimensional
spacetime, $SO(D-1,2)$,\ to those of the worldsheet, $SO(2p,2)$. In a
supersymmetric theory, this would mean a reduction from the SUSY extension
of the first, to a SUSY extension of the latter. This analysis will be the
subject of a forthcoming paper \cite{EGMZ}.

The branes considered here are $\delta $-like objects in the spacetime
manifold. The interpretation of distributions of this sort in General
Relativity is obscure if they appear in the metric, and great care must be
taken to avoid inconsistencies arising from products and inverses of metric
components that enter in Einstein's Equations \cite%
{Geroch-Traschen,Balasin-Nachbagauer,Steinbauer-Vickers,Clarke-Vickers-Wilson}%
. In Chern-Simons theories, this problem does not arise because the field
equations contain only exterior products of forms, which always produce well
defined distributional products.

An additional open question is how to generalize the notion of transgression
for CS forms either of different degree or for intersection of several $p$%
-branes of different dimensionality.

In general, BPS states will \ give rise to fermionic zero modes on the
branes due to the partial breaking of supersymmetry. In contrast with the
standard supergravity, the transformation law for the fields in CS
supergravities is that of the connection, $\delta A=D\lambda $. Therefore,
the fermionic zero modes can be written as $\delta \psi =D\epsilon $, where
the covariant derivative is evaluated on the BPS background retaining only
the generators of the unbroken symmetries. Consequently, as in the standard
case, BPS states give fermionic zero modes for free.

The question remains about the dynamics of branes as effected by their
interaction with the connection field. In order to address this issue, one
should postulate an action principle for free branes and this in turn
requires to decide whether the branes are fundamental objects themselves or
are given as functions of more fundamental matter fields, as for example $%
j\sim \bar{\psi}\Gamma \psi $. In that case, those dynamic source can lead
to spontaneous symmetry breaking or Higgs mechanism, as already mentioned.

\section*{Acknowledgments}

The authors are pleased to thank C. Bunster, J. Edelstein, A. Garbarz, G.
Giribet, C. Mart\'{\i}nez and R. Olea for many enlightening conversations
and suggestions. They would also like to thank M. Ba\~{n}ados, B.
Cuadros-Melgar, A. Gomberoff, M. Henneaux, P. Mora and S. Willison for
useful discussions. This work was supported in part through FONDECYT grants
\#11070146, 1061291, 7070117 and 7080201. The Centro de Estudios Cient\'{\i}%
ficos (CECS) is funded by the Chilean Government through the Millennium
Science Initiative and Conicyt's \textit{Programa de Financiamiento Basal
para Centros de Excelencia}. CECS is also supported by a group of private
companies which at present includes Antofagasta Minerals, Arauco, Empresas
CMPC, Indura, Naviera Ultragas and Telef\'{o}nica del Sur.

\appendix{}

\section{Coupling branes to Abelian CS theories \label{Abelian}}

Here we show a few explicit examples of classical solutions for the coupling
between an Abelian connection and simple external sources.

\subsection{1. Point charge in $2+1$\ dimensions.}

Consider the CS system for a $U(1)$\ connection in $2+1$\ dimensions. The
only brane that can couple to this connection is a $0$-brane (point charge).
In the presence of a point particle sitting at rest, the action reads
\begin{eqnarray}
I_{2+1}[A,j] &=&\kappa \int\limits_{M^{2+1}}\left( \frac{1}{2}%
\,AdA-j_{[0]}\,A\right)  \notag \\
&=&\kappa \left( \frac{1}{2}\int\limits_{M^{2+1}}AdA-q_{0}\int\limits_{%
\Gamma ^{1}}A\right) \,.  \label{3D+P-charge}
\end{eqnarray}%
The resulting field equation is%
\begin{eqnarray}
F &=&j_{[0]}  \notag \\
&=&q_{_{0}}\,\delta (\Sigma _{12})\,d\Omega ^{2}\text{ ,}
\label{Monopole eqn}
\end{eqnarray}%
where $\Sigma _{12}$\ is the spatial section in the rest frame of the
particle. The solution for the connection $A$\ is found by direct
integration of (\ref{Monopole eqn}) on a disc. It reads, modulo gauge
transformations,%
\begin{equation}
A=\frac{q_{_{0}}}{2\pi }\,d\phi _{12}+a(\Gamma ^{1})\text{ ,}
\label{2+1monopole}
\end{equation}%
where $\phi _{12}$ is the angle around the origin in the $\Sigma _{12}$
plane, and $a(x^{0})$ is an arbitrary $1$-form on the worldline. This
configuration corresponds to a point charge that produces a timelike
magnetic flux $\Phi =\int \int F=q_{0}$, concentrated on the worldline. The
curvature $F$ can be obtained directly by differentiating (\ref{2+1monopole}%
) and using the identity \cite{Jackiw90},%
\begin{equation}
d(d\phi )=2\pi \,\delta (\Sigma _{12})\,d\Omega ^{2}\text{ }.  \label{ddphi}
\end{equation}

This source generates a static magnetic field, a monopole in $2+1$
dimensions. The quantization of the magnetic flux, $\kappa F=2n\pi \hbar $,
would be a consequence of requiring that the holonomies around the monopole
be quantum mechanically unobservable. In that case, the allowed values for
the \textquotedblleft magnetic charge\textquotedblright\ must be quantized
by Dirac's rule, $\kappa q_{_{0}}=nh$\ .

\subparagraph{2. Point source in $4+1$\ dimensions.}

The action for a $U(1)$ CS connection in $4+1$ dimensions coupled to a point
source is%
\begin{eqnarray}
I_{4+1}[A,j] &=&\kappa \int\limits_{M^{4+1}}\left( \frac{1}{3}%
\,A(dA)^{2}-j_{[0]}\,A\right)  \notag \\
&=&\kappa \left( \frac{1}{3}\int\limits_{M^{4+1}}A(dA)^{2}-q_{_{0}}\int%
\limits_{\Gamma ^{1}}A\right) \,.
\end{eqnarray}%
The corresponding field equations read%
\begin{eqnarray}
FF &=&j_{[0]}  \label{FF} \\
&=&q_{_{0}}\,\delta \left( T^{4}\right) \,d\Omega ^{4}\,,
\end{eqnarray}%
which have many solutions for $F$, in particular,%
\begin{equation*}
F=\frac{\sqrt{q_{_{0}}}}{2}\,\left[ \delta (\Sigma _{12})d\Omega
_{12}^{2}+\delta (\Sigma _{34})\,d\Omega _{34}^{2}\right] \,.
\end{equation*}%
Permutations of the labels $1,2,3,4$ clearly yield the same result (\ref{FF}%
). Hence, the general solution\ is a linear combination,
\begin{equation}
F=\sqrt{q_{_{0}}}\sum_{i>j=1}^{4}\xi _{ij}\,\delta (\Sigma _{ij})\,d\Omega
_{ij}^{2}\,,\ \qquad \text{where\quad }\ \frac{1}{8}\,\xi _{_{ij}}\xi
_{_{kl}}\,\epsilon ^{ijkl}=1\,,  \label{F-0brane-D=5}
\end{equation}%
and the connection reads%
\begin{equation}
A=\frac{\sqrt{q_{_{0}}}}{2\pi }\,\sum_{i>j=1}^{4}\xi _{ij}\,d\phi _{ij}+%
\mathcal{A}(x^{0})\,,  \label{A-0brane-D=5}
\end{equation}%
where $\mathcal{A}$ is an arbitrary $1$-form. The six arbitrary coefficients
$\xi _{ij}$ reflect the fact that higher dimensional CS theories have larger
degeneracies, which are not found in more standard gauge theories, like in
the Maxwell or Yang-Mills cases.

\subparagraph{3. Two-brane in $4+1$\ dimensions.}

A five-dimensional CS theory for a $U(1)$ connection, can couple to a
two-dimensional membrane. The action now reads,%
\begin{eqnarray}
I_{4+1}[A,j] &=&\kappa \int\limits_{M^{4+1}}\left( \frac{1}{3}%
\,A(dA)^{2}-j_{[2]}\,AdA\right)  \notag \\
&=&\kappa \left( \frac{1}{3}\int\limits_{M^{4+1}}A(dA)^{2}-q_{2}\int%
\limits_{\Gamma ^{3}}AdA\right) \,.
\end{eqnarray}%
The corresponding field equation,
\begin{equation}
F{\small \wedge (}F-j_{[2]})=0,  \label{FF=jF}
\end{equation}%
is degenerate: a portion of solution space is not determined by this
equation. There are four (almost) obvious solutions for any source $j_{[2]}$:

i) The connection is flat everywhere,%
\begin{equation}
F^{(I)}=dA^{(I)}=0\text{ .}  \label{F(1)}
\end{equation}

ii) The curvature is given by the source,%
\begin{equation}
F^{(II)}=j_{[2]}\,.  \label{F(2)}
\end{equation}

Since the current is a $2$-form source on the section transverse to the
worldvolume of the brane,%
\begin{equation}
j_{[2]}=q_{_{2}}\,\delta (\Sigma _{34})\,d\Omega ^{2}\text{ },
\end{equation}%
in the second case, the connection reads%
\begin{equation}
A^{(II)}=\frac{q_{_{2}}}{2\pi }\,d\phi _{34}+\mathcal{A}(\Gamma ^{3})\,,
\end{equation}%
up to gauge transformations. Here $\mathcal{A}$\ is an arbitrary $1$-form
defined on $\Gamma ^{3}$.

iii) A third obvious solution to (\ref{FF=jF}) contains (\ref{F(1)}) and (%
\ref{F(2)}) as particular cases,%
\begin{equation}
F^{(III)}=d\mathcal{A}^{(III)}(T^{2})\text{ ,}  \label{F(3)}
\end{equation}%
where $\mathcal{A}^{(III)}$ is any 1-form defined on the two-dimensional
transverse space $T^{2}$.

iv) The fourth solution, independent from those above, is%
\begin{equation}
F^{(IV)}=\frac{1}{2}\,j_{[2]}+d\mathcal{A}^{(IV)}(\Gamma ^{3}),  \label{F(4)}
\end{equation}%
where $\mathcal{A}^{(IV)}$ is any 1-form on the worldvolume of the brane $%
\Gamma ^{3}$. One could also add a term $\alpha (\Gamma ^{3}){\small \wedge
\beta }(T^{2})$, where $\alpha ${\small \ }and{\small \ }$\beta $ are
1-forms on the worldvolume and in the transverse space, respectively.
However, this would require additional algebraic constraints which could not
be easily implemented. Up to gauge transformations, the connection for (\ref%
{F(4)}) reads%
\begin{equation}
A^{(IV)}=\frac{q_{_{2}}}{4\pi }\,d\phi _{34}+\mathcal{A}^{(IV)}(\Gamma ^{3})%
\text{ },
\end{equation}%
where $\phi _{34}$\ is the angle in the transverse space surrounding the
brane.

The fact that the phase space contains various disconnected sectors is a
general feature, related to the presence of degeneracies in the symplectic
form and to the existence of irregular constraint structures in
higher-dimensional CS theories. These issues have been analyzed in \cite%
{Miskovic-Zanelli,Saavedra-Troncoso-Zanelli,Miskovic-Troncoso-Zanelli}.

\subparagraph{4. Generalization to higher dimensions.}

The general recipe for a co-dimension $2$ brane --or $(2n-2)$-brane-- in $%
D=2n+1$ dimensions can be easily presented. The transverse space is $2$%
-dimensional ($T^{2}$), as in the examples discussed above ($D=5$, $n=2$),
and the field equations read%
\begin{equation}
F^{n}=j_{[n-1]}F^{n-1}\,,
\end{equation}%
which have at least four sectors in the space of solutions (up to gauge),
\begin{equation*}
\begin{array}{ll}
F^{(I)}=0\,, & A^{(I)}=0\,, \\
F^{(II)}=j_{[n-1]}\,, & A^{(II)}=\dfrac{q_{_{n-1}}}{2\pi }\,d\phi _{D-2\text{
}D-1}\text{ }, \\
F^{(III)}=dA^{(III)}(T^{2})\,, & A^{(III)}(T^{2})\text{ arbitrary}\,, \\
F^{(IV)}=\dfrac{1}{n}\,j_{[n-1]}+d\mathcal{A}^{(IV)}(\Gamma )\,,\qquad &
A^{(IV)}=\dfrac{q_{_{n-1}}}{2n\pi }\,d\phi _{D-2\text{ }D-1}+\mathcal{A}%
^{(IV)}(\Gamma )\text{ }.%
\end{array}%
\end{equation*}%
These sectors have different residual symmetries and, correspondingly,
different degrees of freedom for the propagating modes around the
corresponding vacua. Those degrees of freedom are governed by the
\textquotedblleft homogeneous\textquotedblright\ ($q_{n-1}=0$) parts of the
solutions and, as mentioned above, this is a generic feature of
higher-dimensional CS theories.

Similar co-dimension $2$\ defects were also found in even-dimensional
topological field theories: a $6D$\ topological density in a six-dimensional
AdS spacetime with a four-dimensional topological defect, induces an
effective theory on the defect whose dynamics is that of $4D$\ Einstein
gravity \cite{AWZ}.

In the other extreme, the case of $0$-branes in $2n+1$ dimensions, can be
easily generalized as well. The field equations read%
\begin{equation*}
F^{n}=j_{[0]}\text{ },
\end{equation*}%
where $j_{[0]}$ is a $2n$ form on the $2n$-dimensional transverse space to
the worldline of the $0$-brane. The solution takes the form
\begin{equation}
F=\sqrt{q_{_{0}}}\sum_{i>j=1}^{2n}\xi _{ij}\,\delta (\Sigma _{ij})\,d\Omega
_{ij}^{2}\,,\qquad \text{where}\qquad \frac{1}{2^{n}n!}\,\xi
_{_{i_{1}j_{1}}}\cdots \xi _{_{i_{n}j_{n}}}\,\epsilon ^{i_{1}j_{1}\cdots
i_{n}j_{n}}=1\,,
\end{equation}%
and the connection is%
\begin{equation}
A=\frac{\sqrt{q_{_{0}}}}{2\pi }\,\sum_{i>j=1}^{2n}\xi _{ij}\,d\phi _{ij}+%
\mathcal{A}(x^{0})\,.
\end{equation}

The even more radical case of a ($-1$)-brane could be conceived as well. It
would correspond to an object whose worldvolume is zero-dimensional, namely,
a charged instanton \cite{Bunster}.

\section{Regularization of the curvature for the higher-dimensional 0-brane
\label{0 brane regularization}}

Consider the metric (\ref{0-brane metric}) of $D$-dimensional AdS space with
surface deficit $\alpha \Omega ^{D-2}$, where\textbf{\ }$\alpha \in \lbrack
0,1]$\ is the fraction of the topological defect and $\Omega ^{D-2}$ is the
surface area of a unit sphere $S^{D-2}$. For $\alpha >0$, this metric has a
singularity at the origin $r=0$ and its regularization consists in smoothing
out the topological defect $\alpha $ by making the replacement $\alpha
\rightarrow \alpha _{\epsilon }(r)$, where $\alpha _{\epsilon }(r)$\ is
chosen as\textbf{\ }%
\begin{equation}
\alpha _{\epsilon }(r)=\frac{\alpha \,r^{2}}{r^{2}+\epsilon ^{2}}\,.
\label{f}
\end{equation}%
Note that for fixed finite $r$, $\alpha _{\epsilon }(r)\rightarrow \alpha $\
in the limit\ $\epsilon \rightarrow 0$, and for fixed finite\ $\epsilon $, $%
\alpha _{\epsilon }(r)\rightarrow 0\,$in the limit\ $r\rightarrow 0$. The
regularized vielbein can be taken as%
\begin{eqnarray}
e_{\epsilon }^{0} &=&f(r)\,dt\,, \\
e_{\epsilon }^{1} &=&\frac{dr}{f(r)}\,, \\
e_{\epsilon }^{p} &=&\left[ 1-\alpha _{\epsilon }(r)\right] r\,\tilde{e}%
^{p}\,,
\end{eqnarray}%
where $f^{2}=1+\frac{r^{2}}{\ell ^{2}}$ is the AdS metric function and $%
\tilde{e}^{p}$ is the vielbein of the $(D-2)$ unit sphere, $d\Sigma
_{D-2}^{2}=\delta _{pq}\,\tilde{e}^{p}\tilde{e}^{q}$. \ Assuming that
torsion vanishes for $r>0$, the components of the regularized spin
connection are%
\begin{eqnarray}
\omega _{\epsilon }^{01} &=&f\,f^{\prime }\,dt\,, \\
\omega _{\epsilon }^{1p} &=&f\,\left( r\alpha _{\epsilon }\right) ^{\prime
}\,\tilde{e}^{p}\,, \\
\omega _{\epsilon }^{pq} &=&\tilde{\omega}^{pq}\,,
\end{eqnarray}%
that gives the regularized AdS curvature, $F_{\epsilon }^{ab}=R_{\epsilon
}^{ab}+\frac{1}{\ell ^{2}}\,e_{\epsilon }^{a}e_{\epsilon }^{b}$, in the form%
\begin{eqnarray}
F_{\epsilon }^{0p} &=&X_{\epsilon }(r)\,dt\,\tilde{e}^{p}\,,  \label{Fop} \\
F^{1p} &=&Y_{\epsilon }(r)\,dr\,\tilde{e}^{p}\,, \\
F^{pq} &=&Z_{\epsilon }(r)\,\tilde{e}^{p}\tilde{e}^{q}\,.  \label{Fpq}
\end{eqnarray}%
For later use, it is convenient to analyze the behavior of $X_{\epsilon }$, $%
Y_{\epsilon }$ and $Z_{\epsilon }$ for small radius, $r=\epsilon \eta $.
Expanding in powers of $\epsilon $ and finite $\eta $, one finds
\begin{eqnarray}
X_{\epsilon }(\epsilon \eta ) &=&-\frac{2\alpha k\eta ^{3}}{\ell ^{2}\left(
\eta ^{2}+1\right) ^{2}}\,\epsilon +\mathcal{O}(\epsilon ^{3})\,,  \label{X}
\\
Y_{\epsilon }(\epsilon \eta ) &=&\frac{2\alpha \eta \left( 3-\eta
^{2}\right) }{\left( \eta ^{2}+1\right) ^{3}}\,\frac{1}{\epsilon }+\mathcal{O%
}(\epsilon )\,, \\
Z_{\epsilon }(\epsilon \eta ) &=&1-\frac{\left[ \left( 1-\alpha \right) \eta
^{4}+\left( 2-3\alpha \right) \eta ^{2}+1\right] ^{2}}{\left( \eta
^{2}+1\right) ^{4}}+\mathcal{O}(\epsilon ^{2})\,.  \label{Z}
\end{eqnarray}%
The remaining components of $F_{\epsilon }^{AB}$ vanish. Note that the
torsion $F_{\epsilon }^{aD}=(De^{a})_{\epsilon }$ identically vanishes in
this regularization.

We know that $\left( F_{\epsilon }\right) ^{n}$ is singular and, on account
of the field equations (\ref{e.o.m.}) for $0$-branes, and using (\ref{Fop}-%
\ref{Fpq}), the non-vanishing components of the regularized current are%
\begin{equation}
(j_{\epsilon })_{a}=\frac{1}{2^{n}}\,\varepsilon _{aa_{1}b_{1}\cdots
a_{n}b_{n}}\,F_{\epsilon }^{a_{1}b_{1}}\cdots F_{\epsilon }^{a_{n}b_{n}}\,.
\end{equation}%
Moreover, from the properties of the functions $X_{\epsilon },Y_{\epsilon }$
and $Z_{\epsilon }$, it is straightforward to show that the current is%
\begin{equation}
(j_{0})_{\epsilon }=\frac{n}{2^{n-1}}\,\varepsilon _{01p_{1}\cdots
p_{2n-1}}\,Y_{\epsilon }(r)\,Z_{\epsilon }^{n-2}(r)\,dr\tilde{e}^{p_{1}}\,%
\tilde{e}^{p_{2}}\cdots \tilde{e}^{p_{2n-1}}\,.
\end{equation}%
Treating this source as a distribution that is regular everywhere, we
multiply it by a test function $\Psi $ with support on the spatial section
of the space-time. Since the source is spherically symmetric, one can take
the average over the angular part. Denoting the area of a $S^{2n-1}$ sphere
as $\Omega ^{2n-1}=\int \varepsilon _{01p_{1}\cdots p_{2n-1}}\tilde{e}%
^{p_{1}}\cdots \tilde{e}^{p_{2n-2}}$, one obtains%
\begin{equation}
\lim_{\epsilon \rightarrow 0}\int \Psi (\vec{r})\,(j_{0})_{\epsilon
}=\lim_{\epsilon \rightarrow 0}\,\frac{n}{2^{n-1}}\,\Omega
^{2n-1}\int\limits_{0}^{\infty }dr\,\bar{\Psi}(r)Y(r)\,Z^{n-1}(r)\text{ ,}
\end{equation}%
and, after changing the integration parameter $r=\epsilon \eta $\ and using
the properties (\ref{X}-\ref{Z}), one obtains in the limit,%
\begin{equation*}
\lim_{\epsilon \rightarrow 0}\int \Psi (\vec{r})\,(j_{0})_{\epsilon }\equiv
q_{0}^{(2n+1)}(\alpha )\,\bar{\Psi}(0)\,.
\end{equation*}

The charge in $2n+1$ dimensions is given by the polynomial in the
topological defect $\alpha $,
\begin{equation}
q_{0}^{(2n+1)}=\Omega ^{2n-1}\,\frac{n\alpha ^{n}}{2^{n-1}}%
\int\limits_{0}^{\infty }dz\,\frac{z^{n-1}\left( 3-z\right) \left(
3+z\right) ^{n-1}}{\left( z+1\right) ^{4n-1}}\,\left[ \left( 2-\alpha
\right) z^{2}+\left( 4-3\alpha \right) z+2\right] ^{n-1}\,.
\end{equation}%
For example, we have%
\begin{eqnarray}
q_{0}^{(3)} &=&2\pi \alpha \,, \\
q_{0}^{(5)} &=&\frac{4\pi \alpha ^{2}}{3}\,\left( 3-\alpha \right) \,, \\
q_{0}^{(7)} &=&\frac{2\pi ^{2}\alpha ^{3}}{5}\,(\alpha ^{2}-5\alpha +\frac{20%
}{3})\,,
\end{eqnarray}%
in dimensions $3,$ $5$ and $7$, respectively.

\end{document}